\documentstyle[preprint,aps]{revtex}

\begin{document}

\title{
QCD Phase Transition and Hadron Bubble Formation
in the Dual Ginzburg-Landau Theory
} 
\vspace{3.5cm}
\author{ 
H. Ichie, H. Monden$^*$,  H. Suganuma and H. Toki
}
\address{
Research Center for Nuclear Physics (RCNP), 
Osaka University, Ibaraki 567, Osaka, Japan\\
$^*$ Department of Physics, Tokyo Metropolitan University, 
Hachioji 192, Tokyo, Japan}

\maketitle

\vspace{0.9cm}

\begin{abstract}
We study the QCD phase transition at finite temperature
and discuss the hadron bubble formation in the dual 
Ginzburg-Landau theory, which is an effective theory of 
QCD and describes the color confinement via   QCD-monopole condensation.
We formulate the effective potential at various temperatures and find 
that thermal effects reduce the QCD-monopole condensate and 
bring a first-order deconfinement phase transition.
Based on this effective potential at finite temperature,
we investigate  properties of  hadron bubbles created
in the early Universe and discuss the  hadron bubble formation process.  
\end{abstract}

\section{Introduction}

At low temperature and low density, 
colored particles, quarks and gluons are confined in hadrons 
due to the nonperturbative nature of the QCD vacuum.
However, as the temperature increases above  a certain critical temperature,
these colored particles would be liberated like free particles
and the vacuum becomes quark gluon plasma (QGP) phase\cite{greiner}.
This phenomenon is called  QCD phase transition
and many physists try to create QGP
 in  high-energy heavy-ion
collisions (e.g. RHIC at Brookhaven).
After two heavy ions collide, 
and pass through each other, the huge energy deposition at central
region leads to QGP\cite{greiner}.
The QCD phase transition also happened as a real event in the early Universe,
which strongly influenced the afterward
nucleosynthesis.
Thus, the QCD phase transition is interesting in the various fields.

The QCD phase transition is caused by the change of QCD vacuum.
In low energy region, the QCD vacuum is regarded as the dual version of 
superconductor\cite{Nambu}.
Condensation of  the magnetic charge makes the 
color-electric field  excluded from the QCD vacuum, which leads to the
color confinement.
This picture is constructed after abelian gauge fixing as 't Hooft 
proposed\cite{t Hooft}.
The most relevant gauge  for the discussion of confinement is the abelian gauge,
where QCD-monopoles appear as topological objects having the magnetic
charges.
This is modeled in  the dual Ginzburg Landau (DGL) 
theory\cite{maedan,suganuma},
which has a strong connection with  QCD and are supported from the recent
results of lattice QCD\cite{kro}.

\section{Dual Ginzburg-Landau Theory}
The DGL lagrangian\cite{suganuma,ichie}
in  the pure gauge system is written by using
the dual gauge field,  $B_\mu=\vec B_\mu \cdot \vec H = B_\mu^3 T^3+B_\mu^8 
T^8$, and the QCD-monopole field, $\chi=\sum_a \sqrt{2}\chi_a E_a$
($E_1=\frac{1}{\sqrt{2}}(T_6+iT_7),
 E_2=\frac{1}{\sqrt{2}}(T_4-iT_5),
 E_3=\frac{1}{\sqrt{2}}(T_1+iT_2)$);
\begin{eqnarray}
{\cal L}_{DGL}={\rm tr}{\hat {\cal L}} \nonumber 
\end{eqnarray}
\begin{eqnarray}
\hat {\cal L}=-{1 \over 2}
(\partial _\mu  B_\nu -\partial _\nu B_\mu )^2
+[\hat{D}_\mu, \chi]^{\dag}[\hat{D}^\mu, \chi]
-\lambda ( \chi^{\dag} \chi -v^2)^2,
\label{eq:dgllag}
\end{eqnarray}
where $\hat{D}_\mu=\hat{\partial}_\mu +igB_\mu$ is the dual 
covariant derivative.
The dual gauge field $B_\mu$ is defined on the dual gauge manifold
U(1)${}^m_3\times$U(1)${}^m_8$,
which is the dual space of the maximal torus subgroup
U(1)${}^e_3\times$U(1)${}^e_8$
embedded in  the original gauge group SU(3)\cite{suganuma}.
The abelian field strength tensor is written as  
$F_{\mu \nu}=
{}^* (\partial\wedge B)_{\mu\nu}$
so that the role of the electric and magnetic field are interchanged 
in comparison with the ordinary $A_\mu$  description.
The QCD-monopole has magnetic charge  $g\vec \alpha$,
where $\vec \alpha$ is the root vector, and $g$ is the dual gauge 
coupling constant.
If the QCD-monopole condenses, $|\chi_\alpha| =v$,
the dual gauge field  acquires mass
$m_B = \sqrt{3}gv $
by the dual Higgs mechanism and  the color-electric field 
is excluded from the QCD vacuum.
In the DGL theory, QCD-monopole condensate is the order parameter of 
the deconfinement phase transition\cite{ichie}.

\section{QCD Phase Transition}
To investigate the QCD phase transition  at finite temperature,
we formulate the effective potential\cite{ichie}(thermodynamical potential)
as the function of QCD-monopole condensate $|\chi_\alpha|=\bar \chi$,
\begin{eqnarray}
 V_{\rm eff}(\bar \chi ;T) =   3 \lambda ( \bar \chi^2 - v^2 )^2 
          &+ 3 {T \over \pi^2} \int_0^\infty  dk k^2 \ln{
           \left(  1 - e^{ - \sqrt{ k^2 + m_B^2}/T }
           \right)  } \nonumber \\
          &+ {3 \over 2} {T \over \pi^2}\int_0^\infty  dk k^2 \ln{
           \left(  1 - e^{ - \sqrt{ k^2 + m_{\chi}^2}/T }
           \right)  }, 
\label{eq:Vb}
\end{eqnarray}
where $m_B= \sqrt{3} g \bar \chi$ and $m_\chi=2\sqrt{\lambda} \bar \chi$.
The effective potential is plotted in Fig.1 and
the behavior of the QCD-monopole condensate at various 
temperatures is shown in Fig.2. 
As the temperature increases,  
the QCD-monopole condensate in the physical vacuum, 
$\bar \chi _{\rm phys}(T)$,  decreases and
the broken dual gauge symmetry tends to be restored.
A first order phase transition is found at the  
critical temperature, $T_c \simeq 0.49$ GeV. 
This phase transition is regarded as the 
deconfinement phase transition, 
because there is no confining force among colored particles 
in the QCD vacuum with $\bar \chi _{\rm phys}(T)=0$. 
The critical temperature  $T_c$ = 0.49 GeV, however, seems much larger 
than the recent lattice QCD prediction\cite{karsh,rothe}.
We introduce therefore the temperature dependence 
on the parameter $\lambda $ as,
\begin{eqnarray}
  \lambda (T) = \lambda  \left( {T_c - aT \over T_c}  \right), 
\label{eq:Lam}
\end{eqnarray}
according to the asymptotic freedom property of QCD.
The behavior of the QCD-monopole condensate 
is shown in Fig.3 for the case where the parameter $a$ are chosed so as to
provide the critical temperature at $T_c$=0.2GeV. 
We find a weak first order phase transition.

Using this QCD-monopole condensate at various temperatures,
we next consider the string tension of hadron (Fig.4) and the glueball mass
(Fig.5).
As the temperature increases,  the string tension becomes smaller and drops
rapidly near the critical temperature, which  agrees with the lattice simulation 
data\cite{gao}.
This means that the hadron size at finite 
temperature becomes larger as the temperature increases.
Here, we find also the large reduction of the glueball mass\cite{ichie}
near the critical temperature, where the glueball
 excites violently as its mass 
becomes small and it leads to the QCD phase transition.

\section{Hadron Bubble Formation in Early Universe}

Finally, we consider the application of the DGL theory to  big bang.
As Witten proposed\cite{witten},
if the QCD phase transition is  of first order, 
the hadron and the QGP phase should coexist in early Universe.
Such a mixed phase may cause the inhomogeneity of the Universe in the 
baryon number distribution.
This inhomoginity affects  the primordial nucleo-synthesis\cite{kajino}.

As a result of the 1st order phase transition, 
hadron bubbles appear in the QGP phase near the critical temperature.
We now consider how hadron bubbles are formed
in the DGL theory. 
In the supercooling system,
the free energy of the hadron bubble with radius $R$ profile
$\bar \chi(r;R)$ is written 
using the effective potential at finite temperature,
\begin{eqnarray}
E[\bar \chi(r;R)] = 4 \pi \int_0^{\infty}drr^2
\{ 3( \frac{d\bar \chi(r;R)}{dr})^2 + V_{\rm eff}(\bar \chi ;T)\}.
\end{eqnarray}
We use the sine-Gordon kink ansatz for the profile of 
the QCD-monopole condensate,
\begin{eqnarray}
\bar \chi (r;R)
=\bar \chi _H\tan ^{-1}e^{(R-r) / \delta }/ \tan ^{-1}e^{R / \delta },
\end{eqnarray}
where the thickness of the surface $\delta$ is 
determined by the free energy minimum conditions.
The result is shown in Fig.6.
The QCD-monopole condensate $\bar \chi(r;R)$
is connected smoothly between inside and outside the bubble.
The energy density of the hadron bubble is shown in Fig.7.
It is negative inside and positive near the boundary surface.
The total  energy is roughly estimated as the sum of the surface term
(corresponding to the positive region near the surface) and 
the volume term (corresponding to the negative region inside
the bubble).

The energy of the hadron bubble with radius $R$ is shown in Fig.8.
The bubble whose radius is smaller than critical radius $R_c$ collapses.
Only larger bubbles ( $R > R_c$) are found to grow up from 
the energetical argument.
However, the creation of large bubbles is suppressed because of formation
probability.
In the bubble formation process, there exists a large barrier height $h$
of the effective potential and therefore
the creation of large bubbles needs the large energy fluctuation above 
the barrier height.
Such a process is suppressed  because of the thermal dynamical factor
(proportional
to bubble formation rate), ${\rm exp}(-\frac43 \pi R_{\rm c}^3h/T )$.
Thus, the only small bubbles are created 
practically,
although 
its radius should be larger than $R_{\rm c}$ energitically.
The temperature dependence of the critical radius and the bubble formation
rate is shown in Fig.9 and Fig.10 respectively. 
In the temperature region of the supercooling state, i.e, 
$T_{\rm low} < T < T_{\rm c}$, the hadron bubbles are created.
As the temperature decreases, the size of hadron bubble becomes smaller,
but the bubble formation rate becomes larger.

From these results, we can imagine how the QCD phase transition happens in 
the big bang scenario\cite{fuller}.
At the first stage slightly below $T_{\rm c}$, only large bubbles are 
created but  
its rate is quite small.
As temperature is lowered, smaller bubbles are  created with much 
formation rate.
During this process,
the created hadron bubbles expand with radiating shock wave which
reheats QGP phase\cite{fuller}.
Near $T_{\rm low}$ many small bubbles are violently created.
Finally QGP phase is isolated like the bubble\cite{fuller}.
Such an evolution of the hadron bubble 
can be obtained from the numerical simulation using the DGL theory.

\vspace{0.2cm}

We study the QCD phase transition at finite temperature in the DGL theory.
Thermal effect reduces the QCD-monopole condensate, and the QCD vacuum is changed 
into the QGP phase.
According to the reduction of the  QCD-monopole condensate at high temperature,
the string tension becomes smaller,
the hadron size becomes larger and glueball mass becomes smaller.
We apply the DGL theory to the hadron bubble formation in early Universe.
Using the effective potential, we estimate the size of hadron bubble
at various temperatures.

\vspace{0.2cm}

We would like to thank Prof.~Kajino for stimulating discussions on
the hadron bubble formation. 

\vspace{0.2cm}

{\bf Figure Caption}

\noindent Fig.1
The effective potential at various temperatures as a function of 
the QCD-monopole condensate. The minima of $V_{\rm eff}(\bar \chi;T)$
are plotted by  $\times$.

\noindent Fig.2
The QCD-monopole condensate  $\bar \chi_{\rm phys}(T)$ at the  
minimum of the effective potential as a function of  temperature $T$.

\noindent Fig.3
The QCD-monopole condensate $\bar \chi_{\rm phys}(T)$ at the minimum of the
effective potential as a function of the temperature in the case of 
variable $\lambda(T)$, which  reproduces $T_c$=0.2GeV.

\noindent Fig.4
The string tensions $k(T)$ for a constant $\lambda$ and a 
variable $\lambda(T)$ as functions of the temperature $T$.
The lattice QCD results\cite{gao} in the pure gauge
are shown by black dots near and below the critical temperature.

\noindent Fig.5
The masses $m_B$ and $m_\chi$ of the dual gauge field $B_\mu$
and the QCD-monopole 
field $\chi$ as the function of temperature.

\noindent Fig.6
The profile of the QCD-monopole condensate in the hadron bubble.
Outside of bubble, QCD-monopole does not condense (QGP phase),
while inside the bubble, QCD-monopole condenses (hadron phase).

\noindent Fig.7
The energy density of the hadron bubble.
It is negative inside and positive near the boundary surface.

\noindent Fig.8
The total energy of the  bubble is plotted as a
function of the hadron bubble radius$R$.
The total energy is roughly estimated as the sum of the 
volume term and the surface term.

\noindent Fig.9
The critical radius $R_c$, corresponding to maximum of the energy in Fig.8, 
is plotted as a function of  temperature. 

\noindent Fig.10
The dominant factor of the bubble formation rate
$P(R_c;T) \equiv exp(-4\pi R_c^3 h(T)/3T)$ is plotted
as a function of temperature.
As the temperature decreases, the radius of the created hadron bubble 
becomes smaller, but  the bubble formation rate larger.

\end{document}